\documentclass[a4paper,12pt]{article}
\usepackage{geometry}
\usepackage[utf8]{inputenc}
\usepackage[T1]{fontenc}
\pdfoutput=1
\usepackage{amsmath}
\usepackage{amssymb}
\usepackage{amsfonts}
\usepackage{lmodern}
\usepackage{tikz}
\usepackage{pgfplots}
\usepackage{rotating}
\usepackage{xcolor}
\usepackage{graphicx}
\usepackage{tgheros}
\usepackage{helvet}
\usepackage{amssymb}
\usepackage{tikz}
\usepackage{csquotes}
\usepackage[english]{babel}
\usepackage[]{biblatex}
\usepackage{wrapfig}
\usepackage{float}
\usepackage{multicol}
\usepackage{subcaption}
\usepackage{booktabs}
\usepackage{setspace}
\usepackage{blindtext}
\usepackage[margin=1cm]{caption}
\usepackage{longtable}
\usepackage{layouts}
\definecolor{blue}{RGB}{23,67,106}
\definecolor{red}{RGB}{186,65,12}
\definecolor{almond}{rgb}{0.94, 0.87, 0.8}

\usetikzlibrary{matrix,positioning,calc}
\usetikzlibrary{backgrounds}
\usetikzlibrary{intersections}
\usetikzlibrary{calc}
\usetikzlibrary{arrows}
\usetikzlibrary{backgrounds}
\usetikzlibrary{intersections}
\usetikzlibrary{matrix,positioning,calc}
\usetikzlibrary{backgrounds}
\usetikzlibrary{intersections}
\usetikzlibrary{calc}
\usetikzlibrary{arrows}
\usetikzlibrary{shadows}
\usetikzlibrary{backgrounds}
\usetikzlibrary{intersections}
\usetikzlibrary{arrows}

\usepackage{multicol}
\usepackage{subcaption}
\usepackage{booktabs}
\usepackage{setspace}
\usepackage{blindtext}
\usepackage[margin=1cm]{caption}
\usepackage{longtable}
\usepackage{layouts}
\definecolor{blue}{RGB}{15,47,188}
\definecolor{blue2}{RGB}{15,47,222}
\definecolor{red}{RGB}{186,65,12}
\definecolor{almond}{rgb}{0.94, 0.87, 0.8}
\title{A robust measure of skewness using cumulative statistic calculation}
\author{Mario Schlemmer}
\date{}

\begin{document}

\maketitle\thispagestyle{empty}
\begin{abstract} 
An important aspect of the shape of a distribution is the level of asymmetry. Strong asymmetries play a role in many ecosystems and are found in the size and reproductive success of individuals. But the standard third moment coefficient of skewness has the drawback that it is very sensitive to outliers, which can lead to incorrect interpretations. A new metric is introduced that is based on calculating the cumulative statistics of the Lorenz curve framework, but it evaluates the asymmetry of the underlying distribution. The standard requirements for skewness measures which the proposed measure satisfies are briefly described and it is compared to the moment-based measure using the lognormal distribution with and without outliers. The results demonstrate that the proposed measure behaves similarly for 'normal' distributions, but is robust(not overly sensitive) if there are outliers.
\end{abstract} 

\section*{Introduction}	
If both sides of a univariate continuous distribution mirror one another the distribution is symmetric. If the left tail is longer it is called left-skewed, and right-skewed if the right tail is stretched out more. But if there are values that lie outside the overall pattern of the distribution the classical moment coefficient of skewness $b_1$ lacks robustness. Consider the example dataset that contains the length/width ratio of 88 butter clams collected at Puget Sound (Langkamp \& Hull, 2022). The dataset is summarized by the boxplot in Figure 1, without the outlier the sample is normally distributed and symmetric. With the outlier $b_1=1.49$, a value that indicates a distribution that is highly skewed to the right, but if the outlier is removed $b_1=-0.04$. Similarly, $b_1$ would indicate a heavily left-skewed distribution if there would be a very low value in the dataset instead of the high value. 

	 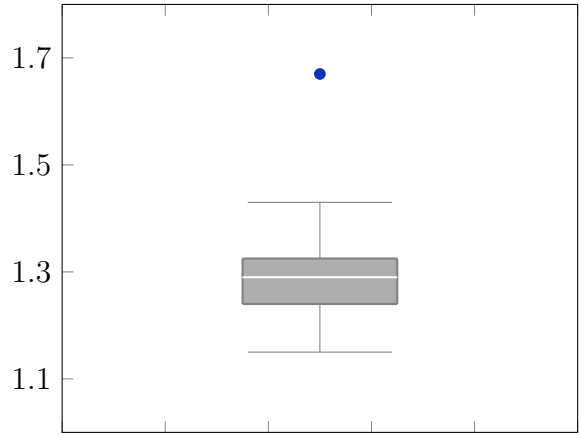
\begin{figure}[h] 	
		\centering
		\begin{tikzpicture}
			\begin{axis}[ xlabel = {},xticklabels={},xtick style={},xticklabel style={xshift=0mm}, xmin =0, xmax = 100,xtick={},  ylabel = {}, ytick={1.1,1.3,1.5,1.7}, ymin =1,ymax = 1.8, ytick pos=left]

				\draw[solid,gray,thick](35,240) -- (65,240);
				\draw[solid,gray,thick](35,325) -- (65,325);
				
				\draw[solid,gray,thick](35,240) -- (35,325);
				\draw[solid,gray,thick](65,240) -- (65,325);
			 \draw[solid,white,thick](35,290) -- (65,290);
			 
			  \draw[solid,gray](50,240) -- (50,150);
			  \draw[solid,gray](50,325) -- (50,430);
			  
			  \draw[solid,gray](36,150) -- (64,150);
			  \draw[solid,gray](36,430) -- (64,430);
			  
			  \node (root0) at (35,240) {}; 
			  \node (lp1) at (65,240) {}; 
			  \node (lp2) at (65,325) {}; 
			  \node (lp3) at (35,325) {};

			  \begin{scope}[on background layer]
			  	\fill[gray!80!white,on background layer,opacity=0.8] (root0.center) -- (lp1.center) --  (lp2.center) -- (lp3.center) -- cycle;
			  	
			  \end{scope}	
			   
				
			\filldraw[blue](50,670)circle(2pt);
				
		\end{axis}	
	
   \end{tikzpicture}
		\caption{\small{Boxplot of the Butter Clams Length/Width Ratio data set, $n=88$.}}
	\end{figure} 
Several statistics that are not affected by outliers have been suggested, among them
the well-known Paerson mode skewness and the closely related medcouple(Brys et al., 2004). The main drawback of these resistant statistics that use the median and quantiles is that they are also not sensitive to extreme tail behavior which can be vital to capturing information about the shape of the distribution. A measure that strikes a balance between tail sensitivity and robustness is based on the Lorenz curve, for which the values are sorted by size and the cumulative proportion of observations(e.g.butter clams) is plotted on the x-axis against the cumulative proportion of measured ratios(y-axis).\\
 \indent This approach provides a graphical representation of inequality among species and individuals that can be used for size, biomass, or fecundity. In economics, it is also used to represent the distribution of assets or the inequality in water and energy consumption. The Lorenz curve coincides with one diagonal of the unit
 square if there is no dispersion in the dataset($p=q$), this diagonal is sometimes called the 45-degree line. If the Lorenz curve is symmetric both sides mirror each other along the other diagonal of the unit square, known as the axis of symmetry(Damgaard, \& Weiner, 2000). But generally, a symmetric Lorenz curve does not imply that the underlying distribution of values is symmetric. It has been pointed out that the most useful content of the Lorenz curve is its distance, $p-q$ from the 45-degree line(Clementi et al., 2019).

\section*{Cumulative Skew}	
 We assume that the sample of a univariate dataset consists of $n$ independent observations $X_n={x_1,x_2,x_3,...,x_n}$, say the size of individuals, the cumulative proportion of individuals is denoted as $p_i$,  the cumulative size as $q_i$. The vertical distances $D_i={d_1,d_2,d_3,...,d_{n-1}}$ between the Lorenz curve and the 45-degree line are calculated by subtracting each $q_i$ from the corresponding $p_i$. If the distribution of $X_n$ is symmetric, then distances at opposite sides of the curve match one another, such a case is presented in Figure 2 which also shows that if the number of distances is uneven, then the distance in the middle has no counterpart. To obtain a measure of skewness positive and negative weights denoted $w_i$ are attached to the distances that lie above and below the median of $p_i$, respectively. No weight is attached to the distance that corresponds to the median. To standardize the sum of weighted distances it is divided by the sum of unweighed distances. 
\begin{figure}[H] 	
	\centering
	\begin{tikzpicture}
		\begin{axis}[scale only axis, xlabel = {Cumulative percentage of individuals}, xmin = 0, xmax = 1, ylabel = {Cumulative percentage of biomass},  ymax = 1, ymin = 0, ytick pos=left]

            \draw[thick,darkgray](5,95) -- (17,95);
            \draw[thick,darkgray](17,95) -- (17,90);
            \draw[thick,darkgray](17,90) -- (5,90);
            \draw[thick,darkgray](5,90) -- (5,95);
            
            \node (roota) at (5,95) {};
            \node (a1) at (5,90) {}; 
            \node (a2) at (11,90) {}; 
            \node (a3) at (11,95) {}; 
           \node (rootc) at (11,95) {};
           \node (a4) at (11,90) {}; 
           \node (a5) at (17,90) {}; 
           \node (a6) at (17,95) {};

            \begin{scope}[on background layer]
            	\path[fill, left color=gray, right color=white!90] (roota.center) -- (a1.center) --  (a2.center) -- 
            	(a3.center)-- (roota.center);
            	
            \end{scope}	
            \begin{scope}[on background layer]
            	\path[fill, left color=white!30, right color=gray] (rootc.center) -- (a4.center) --  (a5.center) -- 
            	(a6.center)-- (rootc.center);
            	
            \end{scope}

            \draw[thick,dashed,red](5,85) -- (17,85); 
            \draw[thick,dashed,green](5,80) -- (17,80);
            \draw[thick,dashed,lightgray](5,75) -- (17,75);
            
            \draw(17,85) node[right] {\footnotesize{$w_i<0$}};
            \draw(17,80) node[right] {\footnotesize{$w_i>0$}};
            \draw(17,75) node[right] {\footnotesize{$w_i=0$}};
	     	\draw(17,92) node[right] {\footnotesize{absolute value of \emph{$w_i$}}};
			\draw(81,71) node[above] {\footnotesize{\rotatebox{40}{45--degree line}}};

			\draw[thick,gray](0,0) -- (100,100);
			\draw[thick, dashed,red](10,10) -- (10,2);
			\draw[thick, dashed,red](20,20) -- (20,5);
			\draw[thick, dashed,red](30,30) -- (30,9);
			\draw[thick, dashed,red](40,40) -- (40,15);
			\draw[thick, dashed,lightgray](50,50) -- (50,23);	
			\draw[thick, dashed,green](60,60) -- (60,35);
			\draw[thick, dashed,green](70,70) -- (70,49);
			\draw[thick, dashed,green](80,80) -- (80,65);
			\draw[thick, dashed,green](90,90) -- (90,82);
			
			\draw[gray](0,0) -- (10,2);
			\draw[gray](10,2) -- (20,5);
			\draw[gray](20,5) -- (30,9);
			\draw[gray](30,9) -- (40,15);	
			\draw[gray](40,15) -- (50,23);
			\draw[gray](50,23) -- (60,35);
			\draw[gray](60,35) -- (70,49);
			\draw[gray](70,49) -- (80,65);
			\draw[gray](80,65) -- (90,82);	
			\draw[gray](90,82) -- (100,100);
			
		\node (root0) at (0,0) {}; 
		\node (lp1) at (10,2) {}; 
		\node (lp2) at (20,5) {}; 
		\node (lp3) at (30,9) {}; 
		\node (lp4) at (40,15) {}; 
		\node (root1) at (100,100) {};
		\node (lp5) at (50,23) {}; 
		\node (lp6) at (60,35) {}; 
		\node (lp7) at (70,49) {}; 
		\node (lp8) at (80,65) {}; 
		\node (lp9) at (90,82) {};

			\begin{scope}[on background layer]
				\path[fill, left color=gray, right color=white!90] (root0.center) -- (lp1.center) --  (lp2.center) -- 
				(lp3.center) -- (lp4.center) -- 
				(lp5.center) -- (50,50) -- cycle;
				
			\end{scope}	
	\begin{scope}[on background layer]
		\path[fill, left color=white!90, right color=gray]  (50,50)--(lp5.center) --  (lp6.center) -- 
		(lp7.center) -- (lp8.center) -- 
		(lp9.center) -- (root1.center) -- cycle;
		
	\end{scope}	
			
		\end{axis}
		
	\end{tikzpicture}
	\caption{Distances between the Lorenz curve and the 45-degree line for a symmetric distribution.} 
\end{figure}
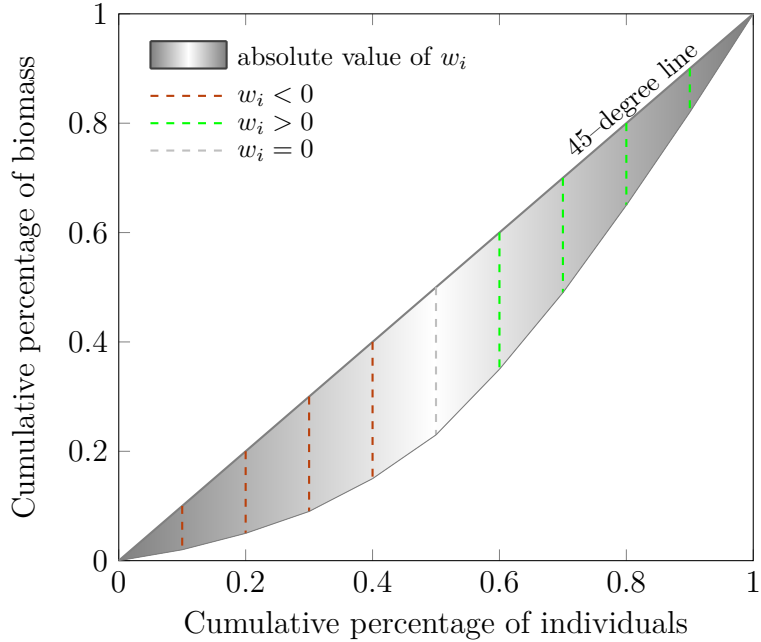
If $r_i$ denotes the ranks of values in the cumulative distribution function, then cumulative skew($CS$)is given by: 
\begin{align}
	\begin{split}\nonumber
		\text{CS}=\frac{\sum (D_i \times w_i)}{\sum D_i} \text{\small{ ,\ where}}
	\end{split}\\
	\begin{split}\nonumber
			w_i=(2\times r_i - n)\times3/n\nonumber
	\end{split} 
\end{align}	
 The weights are bounded by [-3, 3] and $CS$ by[ -1, 1]. For finite $n$ the skewness measure is bounded by [$-1+2/n, 1-2/n]$.\\
\indent The proposed measure $CS$ meets the four requirements for an appropriate skewness measure that have been described in the literature(Groeneveld \& Meeden, 1984; Brys et al.,2004). These properties are:

\noindent \textbf{Property 1:} $CS$ is scale and location invariant, i.e.
\begin{align}\nonumber
		\begin{split}
			F = a\ \times\ F\ +\ b,\ \text{for constants} \ a \ > \ 0 \ \text{and} \ b
		\end{split}  
	\end{align}
  \noindent \textbf{Property 2:} Inverting an asymmetric distribution changes the sign of the skewness.
  \begin{align}\nonumber
  	\begin{split}
  		F(x\times \ -1) = -F(x)
  	\end{split}  
  \end{align}
  These two properties imply Property 3:\\
  \\
  \noindent\textbf{Property 3:} \text{For a symmetric distribution} $\ F \ = \ 0.$\\
  \\
   \noindent \textbf{Property 4:} If $F <_c G$, then $CS(F) \leq CS(G)$\\
   $CS$ respects the c-ordering as defined by van Zwet(1964). A class of distributions that satisfy this c-ordering are Tukey's g-distributions which are calculated from values that are drawn from a normal distribution(Brys et al., 2004; Hoaglin et al, 2011). I calculated g-distributions $G_g$ with different parameters $g$($G_{0.1}$,$G_{0.2}$,...,$G_{1.5}$) and found $G_{g1}$ to c-precede $G_{g2}$ for any $g_1 < g_2$. In Figure 3 I plotted the skewness for g-distributions with different values for parameter $g$. Skewness is increasing faster if there is more variability in the underlying normal distribution, but both lines show a monotonic increase of measured skewness.
  \begin{figure}[H] 	
  	\centering
  	\begin{tikzpicture}
  	\begin{axis}[ xlabel = {\small g}, xmin = -0.1, xmax = 1.6,xtick={0.0,0.5,1,1.5},  ylabel = {\small Skewness},  ymax = 0.9999,ytick={0.0,0.2,0.4,0.6,0.8}, ymin = -0.1, ytick pos=left]
  		\draw[very thick,orange](7,1010)--(25,1010);
  		\draw[very thick,blue](7,940) --(25,940);
  		\draw(23,1000) node[right]{\footnotesize{$M=0, SD=1$}};
  		\draw(23,930) node[right]{\footnotesize{$M=0, SD=3$}};

  		\draw[thick,orange](20,167) --(30,211);
  		\draw[thick,orange](30,211)--(40,255);
  		\draw[thick,orange](40,255) --(50,300);
  		\draw[thick,orange](50,300)--(60,343);
  		\draw[thick,orange](60,343) -- (70,386);
  		\draw[thick,orange](70,386)-- (80,429);
  		\draw[thick,orange](80,429)-- (90,471);
  		\draw[thick,orange](90,471)-- (100,513);
  		\draw[thick,orange](100,513)-- (110,553);
  		\draw[thick,orange](110,553)-- (120,594);
  		\draw[thick,orange](120,594)-- (130,632);
  		\draw[thick,orange](130,632)-- (140,669);
  		\draw[thick,orange](140,669)-- (150,704);
  		\draw[thick,orange](150,704)-- (160,737);

  		\draw[thick,blue2](20,223.9) --(30,346.8);
  		\draw[thick,blue2](30,346.8)--(40,462.2);
  		\draw[thick,blue2](40,462.2) --(50,567.7);
  		\draw[thick,blue2](50,567.7)--(60,661.2);
  		\draw[thick,blue2](60,661.2) -- (70,742.1);
  		\draw[thick,blue2](70,742.1)-- (80,810.1);
  		\draw[thick,blue2](80,810.1)-- (90,866.2);
  		\draw[thick,blue2](90,866.2)-- (100,911.6);
  		\draw[thick,blue2](100,911.6)-- (110,947.5);
  		\draw[thick,blue2](110,947.5)-- (120,976.5);
  		\draw[thick,blue2](120,976.5)-- (130,999);
  		\draw[thick,blue2](130,999)-- (140,1016.5);
  		\draw[thick,blue2](140,1016.5)-- (150,1030.2);
  		\draw[thick,blue2](150,1030.2)-- (160,1040.9);
  		
  	\end{axis}
  	
  \end{tikzpicture}
  \caption{Monotone increasing values of $CS$ as a function of $g$.} 
\end{figure}
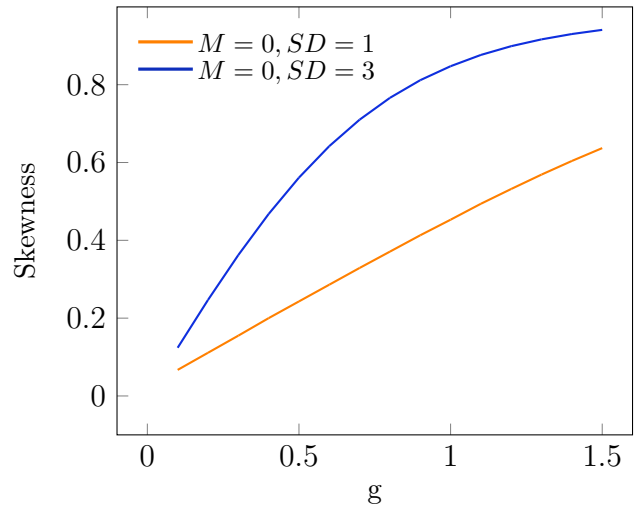

 \section*{Robustness} 
\indent Let us first consider symmetric distributions, property 3 ensures that the probability of negative and positive values is the same if the values are drawn from a distribution that has no bias. I analyzed the behavior for $100.000$ samples of size $n=100$ from a normal distribution and the mean skewness was $-0.000925691$ with a standard error of $0.000135877$. For the heavy-tailed Cauchy distribution the mean was $0.0029298121$ with a standard error of $0.001287708$.\\
 \indent The proposed skewness measure $CS$ is more robust against outliers than the conventional measure $b_1$. For the complete butter clam dataset presented in Figure 1 $CS=0.06$, and when the obvious outlier is removed $CS=-0.02$. To compare the robustness of $CS$ with that of the conventional measure $b_1$ the average skewness for 10.000 random samples of size $n=200$ was calculated for lognormal distributions with different parameter $\sigma$, in addition to two conditions that contaminated lognormal distributions with weak and strong outliers. The lognormal distribution is widely used in ecology to model non-Gaussian dynamics like population abundance and reproductive success of individuals. It is the probability distribution of a random variable whose logarithm is normally distributed. A crucial property of the lognormal distribution is that it is close to a normal distribution for small values of the shape parameter $\sigma$ and gets more right-skewed as $\sigma$ increases.
 
  \begin{center}
  	\begin{centering} 	
  		\begin{tabular}[H]{p{2cm}p{2cm}p{2cm}p{2cm}}
  			\hline
  			\multicolumn{2}{l}{condition}&\multicolumn{1}{r}{$b_1(ave)$}&\multicolumn{1}{r}{$CS(ave)$}\\
  			\hline
  			\multicolumn{2}{l}{1. $\sigma=0.2$}&\multicolumn{1}{r}{0.586}& \multicolumn{1}{c}{0.104}\\
  			\hline
  			\multicolumn{2}{l}{2. $\sigma=0.5$}& \multicolumn{1}{r}{1.598}&\multicolumn{1}{c}{0.241} \\
  			\hline
  			\multicolumn{2}{l}{3. $\sigma=1.0$}&\multicolumn{1}{r}{3.724}&\multicolumn{1}{c}{0.448}\\
  			\hline
  			\multicolumn{2}{l}{4. $\sigma=2.0$}&\multicolumn{1}{r}{7.635}& \multicolumn{1}{c}{0.741}\\
  			\hline
  			\multicolumn{2}{l}{5. $\sigma=0.5$ outliers(high)}&\multicolumn{1}{r}{4.589}&\multicolumn{1}{c}{0.385}\\
  			\hline
  			\multicolumn{2}{l}{6. $\sigma=1.0$ outliers(low)}&\multicolumn{1}{r}{$-1.275$}& \multicolumn{1}{c}{0.153}\\
  			\hline
  			\multicolumn{4}{l}{Table 1. Evaluation of skewness.} 
  		\end{tabular}
  	\end{centering} 
  \end{center}
  
   In Table 1 we see that both $CS$ and $b1$ behave similarly in the first four conditions. As expected, distributions get more and more right-skewed with larger $\sigma$. Condition 5 simulated the effect of 1 to 5 large values in the distribution(0.5\% to 2.5\% of the sample). Here the value of $b_1$ is much larger than in condition 2 which has the same $\sigma$ but no outliers. The effect of the contamination on $CS$ is more moderate. In condition 6 the outliers were 1 to 5 low values. Here the effect of outliers on the average value of $b_1$ is so strong that it now indicates that the distributions are left-skewed, although the vast majority of values were sampled from the heavily right-skewed lognormal distribution ($\sigma=1$). The new measure is again more robust if there are values that lie outside the overall pattern of the distribution.  The strongest differences between the two measures for this sample size were observed if there is only a single strong outlier. This robustness against outliers together with the fulfillment of the four requirements described above and the intuitive appeal through its connection with the Lorenz framework suggests that the new measure $CS$ is a viable alternative to existing measures of skewness.
    
\newpage
\section*{References}
\normalsize{Brys, G., Hubert, M., \& Struyf, A. (2004). A robust measure of skewness. \textit{Journal of}\par\textit{Computationaland Graphical Statistics, 13}(4), 996-1017.}\\
\normalsize{Clementi, F., Gallegati, M., Gianmoena, L., Landini, S., \& Stiglitz, J. E. (2019). Mis-\par measurement of inequality: a critical reflection and new insights. \textit{Journal of Eco-} \par \textit{nomic Interaction and Coordination, 14}(4), 891-921.}\\
\normalsize{Damgaard, C., \& Weiner, J. (2000). \textit{Describing inequality in plant size or fecundity.}\par\textit{Ecology, 81}(4), 1139-1142.}\\
\normalsize{Groeneveld, R. A., \& Meeden, G. (1984). Measuring skewness and kurtosis. \textit{Journal}\par\textit{of the Royal Statistical Society: Series D (The Statistician), 33}(4), 391-399.}\\
\normalsize{Hoaglin, D. C., Mosteller, F., \& Tukey, J. W. (Eds.). (2011). \textit{Exploring data tables,}\par\textit{ trends, and shapes.} John Wiley \& Sons.}\\
\normalsize{Langkamp, G. \& Hull, J.(2022). QELP Data Set 002. [online] Seattlecentral.edu.\par Available at: https://seattlecentral.edu/qelp/sets/002/002.html [Accessed 26 \par September 2022]}\\
\normalsize{Zwet, W.R. van (1964). \textit{Convex Transformations of Random Variables}, Mathematisch\par
	Centrum: Amsterdam.}\\

\end{document}